\documentclass[11pt,a4paper]{article}
\usepackage{amsfonts}
\usepackage{graphicx}
\usepackage{amsmath}
\usepackage{hyperref}
\usepackage{enumerate}
\usepackage{amsmath,amssymb}
\usepackage{slashed,mathrsfs}
\usepackage{feynmp,subfigure}
\usepackage{verbatim,graphicx}
\usepackage[sub,ovp]{psfragx}
\usepackage{overpic}

\RequirePackage{mathrsfs} \RequirePackage[sc]{mathpazo}
\RequirePackage{wasysym} \RequirePackage{setspace}
\newtheorem{theorem}{Theorem}
\newtheorem{acknowledgement}[theorem]{Acknowledgement}

\newcommand{\be}{\begin{equation}}
\newcommand{\ee}{\end{equation}}
\newcommand{\bea}{\begin{eqnarray}}
\newcommand{\eea}{\end{eqnarray}}
\input{tcilatex}
\begin{document}

\date{}
\title{ \rightline{\mbox{\small
{LPHE-MS/10-2016}}}\textbf{A Quantum Approach to Gravitational Waves}}
\author{Salah Eddine Ennadifi$^{1}$\thanks{%
ennadifis@gmail.com} \\
%EndAName
\\
{\small $^{1}$LHEP-MS, Faculty of Science, Mohamed V University, Rabat,
Morocco }\\
}
\maketitle

\begin{abstract}
In light of the Gravitational Waves observation and the discrete structure
of spacetime at high scales $M_{QG}$, we investigate the behavior of
Gravitational Waves and the spacetime deformation patterns in terms of
gravitons and derive the effective results with respect to photons. Using
the $GW150914$ data and related signals, we probe the underlying quantum
scale $M_{QG}\sim 10^{6}GeV$.


\end{abstract}

\newpage

\section{\textit{Introduction}}

The existence of gravitational field, satifying Einstein's equations of
General Relativity (GR), as well as Gravitational Waves (GWs) traveling at
the speed of light is widely agreed \cite{1,2}. Recently, after the
discovery of (GWs) announced by the LIGO and VIRGO detectors in the event
GW150914 \cite{3}, a new era in physics such as tests of GR, heavy black
holes and measurements of astrophysical processes that have been
inaccessible to observations with electromagnetic waves is now open \cite%
{4,5,6,7,8}. In particular, the LIGO detectors have observed GWs from the
merger of two stellar-mass black holes. The detected waveform matches the
predictions of GR for the inspiral and merger of a pair of black holes and
the ringdown of the resulting single black hole. These observations support
the existence of binary stellar-mass black hole systems. This is the first
direct detection of GWs and the first observation of a binary black hole
merger. Fundamentally, these distortions of the neighbouring spacetime may
shed light on the fundamental laws governing our universe such as quantum
aspects of gravity and related models. In particular, an upper limit on the
graviton mass $m_{g}$ $<$ $10^{-22}$ $eV$ has been reported by the LIGO
Collaboration \cite{7}. Gravitons are the field quanta of the gravitational
field, in analogy with photons being the quanta of the electromagnetic
field; they are the ultimate constituents of GWs. New studies of graviton
propagation has been encouraged by the recent thrust\ of interest for GWs in
which the single-quantum behaviour play a central role \cite{7,9,10,11,12}.
With these continuous efforts, especially theoretical, the search for a
quantum description of gravity has rapidly grown \cite{13,14}, and there is
currently a large interest in experimental tests hoping to derive related
effects or unconventional spacetime structure \cite{16,16,17}. Such quantum
signatures of spacetime could be observable only in high energy interactions
over the actual accelerator scales or in very long baseline experiments.

In this work, we are concerned with the question whether we can study the
single-quantum of the GWs, or in other words, the propagation of the
individual gravitons in the fundamental spacetime fabric.We consider for
that the effective discrete spacetime structure at quantum scales $\sim
M_{Planck}$ in terms gravitons of to determine the spacetime deformation
patterns and the graviton propagation behaviour within the possible
dispersive effect. Seen that such effect would be naturally very small and
thus only observable at long baselines experiments, we refer to the $%
GW150914 $ data and the related signals to derive the propagation behavior
and the underlying mass scale $M_{QG}$.

\section{\textit{Classical description}}

According to GR the spacetime geometry is determined by the energy matter
distribution, and the curvature of the spacetime metric is linked to the
energy content of the universe by the equation \footnote{%
Here for simplicity we work in natural units with $\hbar =c=1$.}

\begin{equation}
G_{\mu \nu }=R_{\mu \nu }-\frac{1}{2}Rg_{\mu \nu }=8\pi GT_{\mu \nu }\text{
\  \ }  \label{eq1}
\end{equation}%
where $G_{\mu \nu }$ is the Einstein tensor describing the geometry of the
4D spacetime as the sum of the Ricci tensor $R_{\mu \nu }$ and the metric
tensor $g_{\mu \nu }$, and $T_{\mu \nu }$ is the stress-energy tensor
representing the source of the GWs. The metric $g_{\mu \nu }$ describing the
geometry of spacetime links the spacetime coordinate $dx^{\mu }$ to the
spacetime interval $d\ell ^{2}$ through the relation

\begin{equation}
d\ell ^{2}=g_{\mu \nu }dx^{\mu }dx^{\nu }.\text{\ }  \label{eq2}
\end{equation}%
By the fact that GWs will always be very weak at the earth, the background
curvature can be ignored and the metric can be approximated as that of the
Minkowski flat metric $\eta _{\mu \nu }$, and the equation (\ref{eq1}) can
be linearized in the case of small perturbations. These perturbations can
then be expressed by an approximation as

\begin{equation}
g_{\mu \nu }=\eta _{\mu \nu }+h_{\mu \nu }  \label{eq3}
\end{equation}%
where $\left \vert h_{\mu \nu }\right \vert \ll 1$ is the small perturbation
induced by the GWs. The small changes $\delta \ell ^{2}$ in the spacetime
interval $d\ell ^{2}$ reads

\begin{equation}
\delta \ell ^{2}=h_{\mu \nu }dx^{\mu }dx^{\nu }  \label{eq4}
\end{equation}%
with $\pm d\ell ^{2}$ is the small changes, i.e., contractions and
dillatations in the length $\ell $ of the spacetime fabric. In the weak
field limit and with the most useful gauge choice \footnote{%
Where there is no stress-energy source term $T=0$ and with the most useful
gauge i.e., the TT gauge in which the coordinates are defined by the
geodesics of freely falling test bodies.}, the solution of equation (\ref%
{eq1}) wherein the equation of GR becomes a system of linear equations,
specifically a system of wave equations \cite{17}, corresponds to a 3D wave
equation traveling at the speed of light $c=1$. In the sinusoidal case and
with the symmetry of $h_{\mu \nu }$, the physical part of these waves can be
expressed here in the $z$-direction by the equation

\begin{equation}
h_{\mu \nu }=\delta \ell \epsilon _{\mu \nu }\cos \left( \omega t-kz\right)
\label{eq5}
\end{equation}%
where $\epsilon _{\mu \nu }$\ are the so-called unit polarization tensors $%
\epsilon _{\mu \nu }^{+}$, $\epsilon _{\mu \nu }^{\times }$ with the signs $%
+ $, $\times $ mean that there are just two possible independent
polarization states such as

\begin{equation}
\delta \ell \epsilon _{\mu \nu }=\delta \ell ^{+,\times }  \label{eq6}
\end{equation}%
are the strain amplitudes of each polarization, and $\omega $ and $k$ being
the angular frequency and the wave vector respectively. These classical GWs
have now been confirmed by the recent successful detection in the LIGO
experiment. After this confirmation, the hope for the detection of an
individual quantum of gravity is now desired.

\section{\textit{Quantum approach}}

Theoretically, the quantum picture of GWs is accepted because although any
quantum field theory that describes gravity is non-renormalizable, the
energy scale at which we expect quantum gravity effect to be detectable is
extremely large $\sim M_{Planck}$. Hence, an effective theory valid at low
energies can be used to describe GWs in terms of particles, even if the
exact theory of gravity valid at an arbitrary scale is somewhat different
conceptually. In this context, at the scale where quantum gravitational
effects are expected to be felt $M_{QG}\sim M_{Planck}$, the classical
perception of the spacetime continuum must be altered by a granular
structure. As a result, the vacuum will act as a non trivial medium
caracterized by gravitons of mass $\sim m_{g}$ spaced by elementary length $%
\ell _{QG}\geqslant \ell _{Planck}$ as pictured in the figure 1.

\begin{figure}[th]
\begin{center}
{\includegraphics[width=6cm]{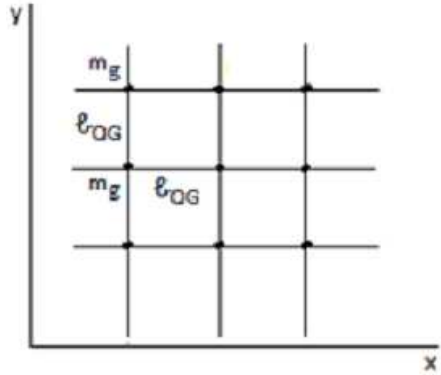}}
\end{center}
\caption{2D discrete spacetime at quantum scale $M_{QG}$.}
\end{figure}

According to this discretisation, the propagating spacetime perturbation (%
\ref{eq5}) can be viewed as coherent superposition of a large number of
gravitons spaced by

\begin{equation}
\ell _{QG}\pm \delta \ell _{QG}\geqslant \ell _{Planck}\text{\ }  \label{eq7}
\end{equation}%
such as $\pm \delta \ell _{QG}$ are now the small perturbations induced by
GWs, i.e., contractions and dillatations of the length $\ell _{QG}$ in the
spacetime fabric. Such deformation patterns are shown in the figure 2 and
the figure 3.

\begin{figure}[th]
\begin{center}
{\includegraphics[width=8cm]{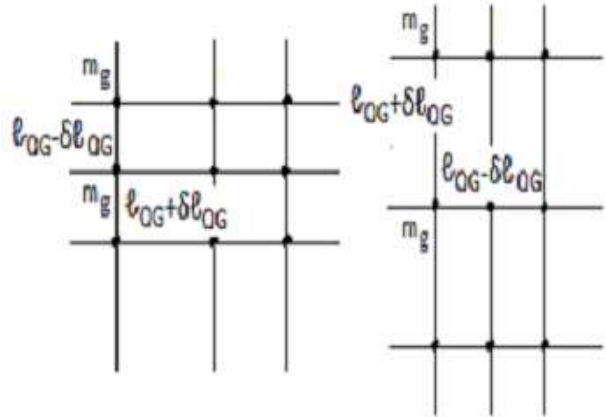}}
\end{center}
\caption{$\protect \delta \ell _{QG}^{+}$ spacetime perturbations at quantum
scale $M_{QG}.$}
\end{figure}

\begin{center}
\begin{figure}[th]
\begin{center}
{\includegraphics[width=8cm]{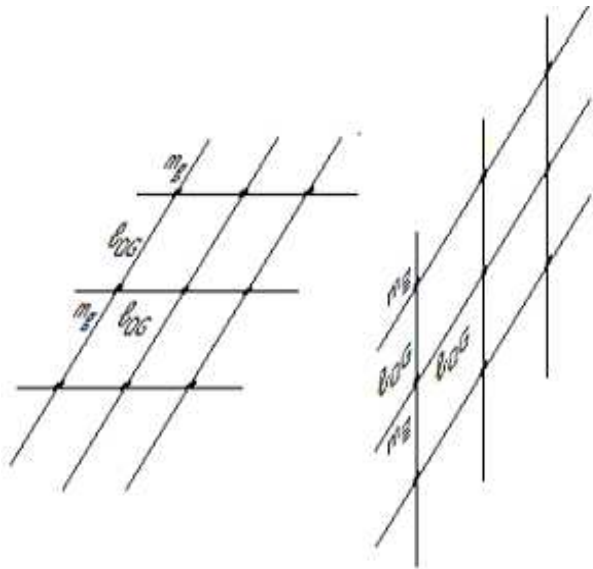}}
\end{center}
\caption{$\protect \delta \ell _{QG}^{\times }$ spacetime perturbations at
quantum scale $M_{QG}.$}
\end{figure}
\end{center}

In flat spacetime (Fig.1) or in a spacetime with small curvature (Fig.2 and
Fig.3) the propagation of gravitons can be described by equations which are
similar to those encountered in electromagnetism. Concretely, in this
quantum description where GWs are viewed as large number of gravitons, a
propagating graviton of energy (frequency) $E_{g}\equiv \omega _{g}$ will be
not immune to thequantum gravity effect, i.e., to the surrounding gravitons.
In genreal, the conservation of energy-momentum during the scattering
process implies that such intercation is of the order

\begin{equation}
\delta \omega _{g}\sim \pm \frac{\omega _{g}^{2}}{M_{QG}}.  \label{eq8}
\end{equation}%
In this direction, the wave equation (\ref{eq5}) can be translated into a
particle equation where the phenomenology can be discussed from a generic
modification of the graviton dispersion relation as

\begin{equation}
\omega _{g}^{2}\simeq k_{g}^{2}+m_{g}^{2}-\frac{\omega _{g}^{4}}{M_{QG}^{2}}
\label{eq9}
\end{equation}%
where we see that the existence of such quntum gravity regime phenomena at a
scale $\sim M_{QG}$ manifests itself through a retardation effect $-\omega
_{g}^{2}/M_{QG}$ in the dynamics of the moving gravitons themselves. At this
stage, it worth mentioning that the effect $\omega _{g}^{2}/M_{QG}$\ in the
particle description (\ref{eq9}) should be physically related somehow to the
perturbation implitude $\delta \ell _{QG}^{+}$ in the wave description (\ref%
{eq9}), since they are already dimensionally related as $\left[ \delta \ell
_{QG}\right] \simeq \left[ \omega _{g}^{2}/M_{QG}\right] ^{-1}$; I beleive
this requiers deep investigations. Here, by neglecting the effect of the
extreme small graviton mass $m_{g}\ll \omega _{g}$ and using the fact that $%
\omega _{g}\  \ll M_{QG}$, straightforward calculations from (\ref{eq9}) lead
to the corresponding effective energy-dependent graviton velocity

\begin{equation}
v_{g}\left( \omega _{g}\right) \simeq \frac{1}{1+\omega _{g}^{2}/M_{QG}^{2}}
\label{eq10}
\end{equation}%
which appear slightly deviated from the velocity of light, and decreases
with increasing energy. In this sens, the higher-energy graviton propagates
slower than the low-energy graviton by the fact that it undergoes a
spacetime attenuation that grows with its energy. This could be then
experienced from (\ref{eq10}), with respect to a light ray propagating with
speed $c=1$ and emitted by the same source at a distance $d$ from the
detector, by the velocity shift

\begin{equation}
\delta v_{\gamma }^{g}\simeq -\frac{\omega _{g}^{2}}{\omega
_{g}^{2}+M_{QG}^{2}}  \label{eq11}
\end{equation}%
and the temporal shift 
\begin{equation}
\  \delta t_{\gamma }^{g}\simeq d\frac{\omega _{g}^{2}}{M_{QG}^{2}}
\label{eq12}
\end{equation}%
being the velocity difference and the time delay of the propagating graviton.

\section{\textit{GW150914} and \textit{results}}

According to $GW150914$ data, the distance estimate to the source is $\sim
10^{9}$ $Ly$, the wave frenquency is $\sim 100$ $Hz$ and the apparently
coincident flash of photons is $>50$ $keV$ observed $0.4$ $s$ later (because
of the scattering process in the intergalactic medium in constrast to
gravitons with their low interaction cross sections) by the Fermi GBM signal 
\cite{18}. The plausibility of the GBM signal has been questioned \cite%
{19,20}. All these data can be summarized, along with the resulting
graviton-photon velocity difference and time delay, in the table 1

\begin{center}
\begin{tabular}{|l|l|}
\hline
Event & $GW150914$ \\ \hline
distance to $GW150914$ : $d_{GW150914}$ & $\sim 10^{9}$ $ly$ \\ \hline
EM frequency: $\omega _{\gamma }\shortmid _{GW150914}$ & $>50$ $keV$ \\ 
\hline
GW frequency: $\omega _{g}\shortmid _{GW150914}$ & $\Game (100)$ $Hz$ \\ 
\hline
time delay: $\delta t_{\gamma }^{g}\shortmid _{GW150914}$ & $\sim 0.4s$ \\ 
\hline
velocity difference: $\delta v_{\gamma }^{g}\shortmid _{GW150914}$ & $%
\lesssim 10^{-17}$ \\ \hline
\end{tabular}

\bigskip 

Table 1: Summary of the $GW150914$ data.
\end{center}

\bigskip Based on the $GW150914$ data (tab. 2) and the fact that $%
M_{QG}\leqslant M_{Planck}$, and according to (\ref{eq11}) and (\ref{eq12}),
we bound the graviton-photon velocity difference and time delay in the table
2

\begin{center}
\begin{tabular}{|l|l|l|}
\hline
Event & time delay: $\delta t_{\gamma }^{g}\shortmid _{GW150914}$ & velocity
difference: $\delta v_{\gamma }^{g}\shortmid _{GW150914}$ \\ \hline
$GW150914$ & $\gtrsim 10^{-27}$ $s$ & $\gtrsim 10^{-52}$ \\ \hline
\end{tabular}

\bigskip 

Table 2: Lower bounds of the $GW150914$ graviton-photon velocity difference
and time delay.
\end{center}

and estimate the underlying quantum gravity scale as

\begin{equation}
M_{QG}\simeq \omega _{g}\shortmid _{GW150914}\sqrt{\frac{d\shortmid
_{GW150914}}{\delta t_{\gamma }^{g}\shortmid _{GW150914}}}\sim 10^{6}GeV.
\label{eq13}
\end{equation}%
Being of the order of $\sim 10^{3}$ $TeV$, the underlying scale $M_{QG}$
appears then to be above the reach of the recent accelerator experiments LHC$%
\sim 10$ $TeV$, but exploration of signatures of the quantum nature of
spacetime remains possible. This unrefined estimate would need to be
precised by a detailed numerical analysis, but it reinforces the point that
mergers of bigger objects can give more powerful constraints on the deep
fabric of spacetime.

\section{\textit{Concluding remarks}}

In this work, a quantum approach to GWs has been proposed based on a
discrete perception of spacetime. Precisely, starting from the classical GR
description and considering a granular structure of spacetime motivated at
high scales $M_{QG}$ in terms of gravitons as its fondamental fabric
(fig.1), a particle descrption of GWs where the spacetime deformation
patterns and the graviton propagation behaviour have been described (\ref%
{eq9}). According to this effective vision, the energy-dependent graviton
velocity along with the corresponding velocity difference and time delay
with respect to the photon have been derived (\ref{eq11}) and (\ref{eq12}).
Then with the analysis of the GW150914 data, lower bounds of these
graviton-photon shifts have been given (tab.2) and the underlying high scale
(\ref{eq13}) has been approached $M_{QG}\sim 10^{6}GeV$. To date,
unfortunately, it is impossible with current technology to observe or
manipulate single gravitons in a lab, so gravitons are still theoretical.
But in spite of the fact that they hav not been observed yet, neutral spin-2
particles as the vectors of the gravitational interaction, are assumed by
most of the quantum gravity models as the one proposed in this work, and
likely massive in contrast to the current dominant opinion as for the case
of neutrinos many years ago. This idea is very speculative, and the model
calculations that we have presented here require refinement. However, it is
well motivated by the basic fact that gravity and spacetime are deeply
related if they are not even one thing.

Finally, we note that if the LIGO Collaboration observes a new signal for a
merger of another pair of black holes \cite{21,22,23}, much greater range of
information will be gathered, useful and deterministic than was possible in $%
GW150914$.

\begin{acknowledgement}
: S-E. Ennadifi would like to deeply thank his family for support.
\end{acknowledgement}

\end{document}